\documentclass[prl,twocolumn,showpacs]{revtex4-1}
\usepackage[latin9]{inputenc}
\usepackage{amsmath}
\usepackage{graphicx}
\usepackage{esint}

\usepackage{color}

\makeatletter

\@ifundefined{textcolor}{}
{%
 \definecolor{BLACK}{gray}{0}
 \definecolor{WHITE}{gray}{1}
 \definecolor{RED}{rgb}{1,0,0}
 \definecolor{GREEN}{rgb}{0,1,0}
 \definecolor{BLUE}{rgb}{0,0,1}
 \definecolor{CYAN}{cmyk}{1,0,0,0}
 \definecolor{MAGENTA}{cmyk}{0,1,0,0}
 \definecolor{YELLOW}{cmyk}{0,0,1,0}
 \definecolor{DARKGREEN}{rgb}{0,0.6,0}
}

\usepackage{bbm}

\newcommand{\Tr}{\textrm{Tr}}
\newcommand{\ket}[2][]{{\mathopen|#2\mathclose\rangle_{#1}}}
\newcommand{\bra}[2][]{{}_{#1}\mathopen\langle #2\mathclose|}
\newcommand{\braket}[3][]{{{}_{#1}\langle#2|#3\rangle_{#1}}}
\newcommand{\proj}[2][]{\ket{#2}_{#1}\bra{#2}}

\makeatother

\begin{document}

\title{Mode engineering for realistic quantum-enhanced interferometry}

\author{Micha{\l} Jachura}

\affiliation{Faculty of Physics, University of Warsaw, Pasteura 5, 02-093 Warsaw, Poland}

\author{Rados{\l}aw Chrapkiewicz}

\email{radekch@fuw.edu.pl}

\affiliation{Faculty of Physics, University of Warsaw, Pasteura 5, 02-093 Warsaw, Poland}

\author{Rafa{\l} Demkowicz-Dobrza\'{n}ski}

\affiliation{Faculty of Physics, University of Warsaw, Pasteura 5, 02-093 Warsaw, Poland}

\author{Wojciech Wasilewski}

\affiliation{Faculty of Physics, University of Warsaw, Pasteura 5, 02-093 Warsaw, Poland}

\author{Konrad Banaszek}

\affiliation{Faculty of Physics, University of Warsaw, Pasteura 5, 02-093 Warsaw, Poland}

\date{\today}


\begin{abstract}
Quantum metrology overcomes standard precision limits by exploiting collective quantum superpositions of physical systems used for sensing, with the prominent example of nonclassical multiphoton states improving interferometric techniques.
Practical quantum-enhanced interferometry is however vulnerable to imperfections such as partial distinguishability of interfering photons. Here we introduce a method where appropriate design of the modal structure of input photons can alleviate deleterious effects caused by another, experimentally inaccessible degree of freedom. This result is accompanied by a laboratory demonstration that a suitable choice of spatial modes combined with position-resolved coincidence detection restores entanglement-enhanced precision in the full operating range of a realistic two-photon Mach-Zehnder interferometer, remarkably around a point which otherwise does not even attain the shot-noise limit due to presence of residual distinguishing information in the spectral degree of freedom. Our method highlights the potential of engineering multimode physical systems in metrologic applications.
\end{abstract}

\maketitle

Quantum phenomena can facilitate and boost the performance of imaging techniques \cite{Brida2010,Ono2013,Schwartz2013,Tsang2009}, sensitive measurements in delicate materials \cite{Wolfgramm2012,Crespi2012}, as well as detection schemes probing subtle physical effects such as gravitational waves \cite{LIGO2013}. These strategies
rely on preparing collective superposition states of multiple probes (for example photons, atoms) to achieve  precision enhancement beyond standard limits \cite{Caves1981,Yurke1986,Bollinger1996,Boto2000, Giovaennetti2006,Paris2008, Banaszek2009,Maccone2011,Toth2014a,Demkowicz-Dobrzanski2014a}. In the optical domain, a common strategy for collective state preparation is to realize multiphoton interference in linear circuits, for example free space or integrated interferometers \cite{Spring2013a,Spagnolo2014,Tichy2014,Motes2014}, fed with nonclassical states of light.
Attainable precision can be however dramatically vulnerable to residual distinguishing information between interfering photons. Standard methods to improve indistinguishability based on filtering are often inadequate, in particular introducing attenuation which may easily diminish the overall benefit of collective state preparation.

The purpose of this paper is to analyze the interplay between degrees of freedom with different experimental accessibility in two-photon interferometry, which is
a canonical example of a quantum-enhanced measurement \cite{Rarity1990}. We demonstrate that detrimental effects caused by distinguishing information present in one degree of freedom that is beyond experimental control or lacks technical means to improve indistinguishability, can be alleviated by mode engineering in another degree of freedom, even though these two remain completely uncorrelated.
This feature is investigated in the case of local phase estimation, whose precision becomes strongly dependent on the operating point if the two photons feeding the interferometer exhibit residual distinguishability. It is shown that a carefully designed preparation and detection scheme for a degree of freedom other than the one causing distinguishability allows one to restore quantum-enhanced precision in the entire operating range of the interferometer. We attribute this effect to non-trivial combination of one- and two-photon interference that turns out to augment phase sensitivity beyond the shot-noise limit. We also present an example indicating that similar enhancement occurs also at higher photon numbers. Prospectively, the results reported here may provide another class of strategies to mitigate effects of imperfections and environmental noise in quantum-enhanced metrology
\cite{Huelga1997,Dorner2008,Knysh2010,Jiang2012,Datta2011,Escher2011,Demkowicz2012}.

\begin{figure}
\includegraphics[width=0.9\columnwidth]{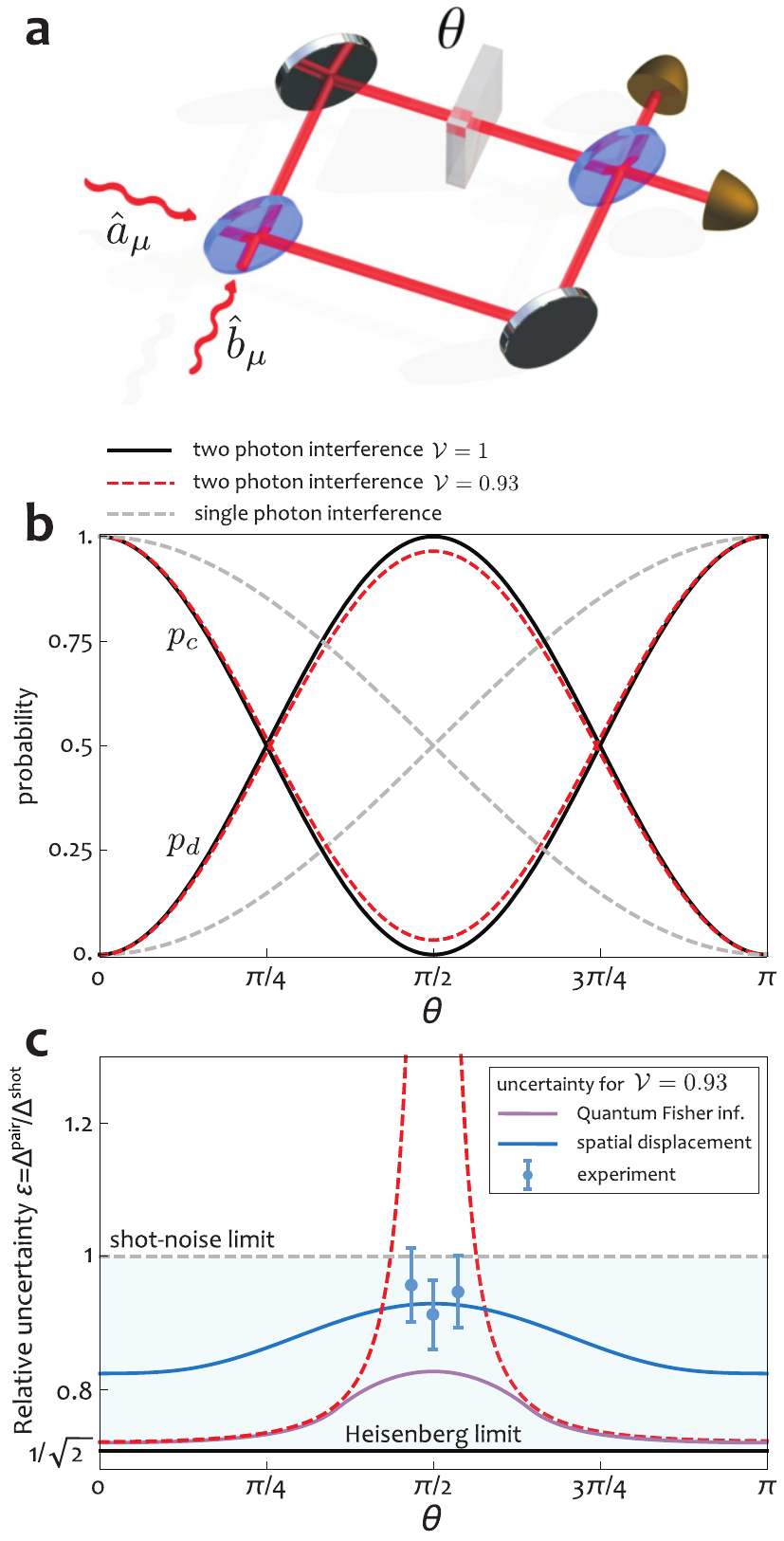}
\protect\caption{{\bf Two-photon interferometer.} (a) In the multimode description, two paths are described by families of annihilation operators $\hat{a}_\mu$ and $\hat{b}_\mu$ subjected pairwise to a unitary map dependent on the phase shift $\theta$. (b) Probabilities of coincidence $p_c$ and double count events $p_d$ at the interferometer output are noticeably affected around $\theta \approx \pi/2$ by imperfect indistinguishability (red dashed lines for the visibility $\mathcal{V}=0.93$) when compared with the ideal case (black solid lines). Grey dashed lines depict standard single-photon interference fringes, when only one input port is illuminated. (c)
Residual distinguishability has a dramatic effect on the phase estimation uncertainty, shown with correspondingly coded lines, which diverges at $\theta = \pi/2$ for non-unit visibility $\mathcal{V}$. Engineering overlap in an additional degree of freedom of the interfering photons and implementing optimal measurement allows one to restore the sub-shot-noise precision over the entire operating range, shown with the purple solid line when the fraction of overlapping pairs is optimized individually for each operating point. The blue solid line depicts an explicit strategy based on introducing a fixed transverse displacement between interferometer inputs and performing spatially resolved detection. These predictions are confirmed by the experimentally determined estimation precision, depicted as solid circles with two-sigma error bars.}
\label{Fig:Fringes}
\end{figure}

The theoretical analysis is complemented with an experiment investigating sensitivity of a balanced Mach-Zehnder interferometer fed with photon pairs. We determine the precision of local phase estimation around the operating point when
the photons coalesce pairwise at the interferometer output ports. In this regime, the residual spectral distinguishability within pairs has a dramatically deleterious effect on the attainable precision. Building on recent advances in spatially resolved single-photon detection  \cite{Peeters2009,Shin2011,Rozema2014,Abouraddy2001,%
Brida2010,Edgar2012,Moreau2014,Lemos2014,Morris2015,Chrapkiewicz2014d,Jachura2015},
we demonstrate that by controlling the input spatial structure
of interfering photons and extracting complete spatial information
at the detection stage it is nevertheless possible to recover the sub-shot-noise
precision. This confirms in proof-of-principle settings
the feasibility of mode engineering techniques for quantum-enhanced interferometry.

\section*{Results}

{\bf Realistic two-photon interferometer.}
A generic two-photon Mach-Zehnder interferometer constructed with a pair
of balanced $50/50$ beam splitters and fed with photon pairs is shown schematically in Fig.~\ref{Fig:Fringes}(a).
The phase shift $\theta$ between the interferometer arms modulates
probabilities of detection events at the output ports which can be
grouped into two types: either the photons exit through different
paths, producing a coincidence event between the detectors monitoring
the ports, or both are found in the same output port leading to a
double event. If the two photons are indistinguishable at the input,
the first beam splitter generates a coherent superposition of both
the photons in one or another arm of the interferometer, which is
the simplest case of a N00N state providing sensitivity that
approaches the Heisenberg limit \cite{Bollinger1996,Boto2000, Giovaennetti2006}.

In the multimode description of the setup one introduces two sets of annihilation operators for upper path modes $\hat{a}_\mu$ and
lower path modes $\hat{b}_\mu$ that are matched pairwise. Individual modes in each arm are mutually orthogonal, i.e.\
$[\hat{a}_\mu, \hat{a}_\nu^\dagger] = [\hat{b}_\mu, \hat{b}_\nu^\dagger] = \delta_{\mu\nu}$.
The unitary map $\hat{U}(\theta)$ implemented by the
interferometer between the input and the output ports transforms pairs of field operators labeled with the same index $\mu$ as
\begin{equation}
\hat{U}^\dagger(\theta)
\begin{pmatrix}
\hat{a}_\mu \\ \hat{b}_\mu
\end{pmatrix}
\hat{U}(\theta)
=
\begin{pmatrix}
\cos{\textstyle\frac{\theta}{2}} & \sin{\textstyle\frac{\theta}{2}}\\
\sin{\textstyle\frac{\theta}{2}} & - \cos{\textstyle\frac{\theta}{2}}
\end{pmatrix}
\begin{pmatrix}
\hat{a}_\mu \\ \hat{b}_\mu
\end{pmatrix}.
\label{Eq:InterfTransformation}
\end{equation}
Partial distinguishability of the interfering photons can
be modeled by assuming that at the input the photon in the upper path occupies a certain mode $\hat{a}_1$, while the lower-path
photon is prepared in a combination of a matching mode $\hat{b}_1$ and another orthogonal mode $\hat{b}_2$ with relative weights ${\cal V}$ and $1-{\cal V}$, where ${\cal V}$ is the visibility parameter specifying the fraction of indistinguishable pairs.
To keep the notation concise we will write the complete two-photon state as pure
\begin{equation}
\label{Eq:PsiInput}
\ket{\psi} = \hat{a}^\dagger_{1} \bigl(\sqrt{\cal V} \hat{b}^\dagger_{1} + \sqrt{1-{\cal V}} \hat{b}^\dagger_{2} \bigr) \ket{\text{vac}},
\end{equation}
and trace the final formulas over the index $\mu=1,2$ for the initial modes $\hat{b}_\mu$. This is equivalent to taking from the start a reduced density matrix
${\cal V} \hat{a}^\dagger_{1} \hat{b}^\dagger_{1} \proj{\text{vac}} \hat{a}_{1} \hat{b}_{1}
+ (1-{\cal V}) \hat{a}^\dagger_{1} \hat{b}^\dagger_{2} \proj{\text{vac}} \hat{a}_{1} \hat{b}_{2} $.
Here $\ket{\text{vac}}$ is the vacuum state of the entire multimode electromagnetic field satisfying $\hat{a}_\mu \ket{\text{vac}}= \hat{b}_\mu \ket{\text{vac}} = 0$ for any $\mu$.
Note that the above model includes the case of partly overlapping wavepackets constructed from a continuum of modes, wherein $\hat{b}_1$ and $\hat{b}_2$ can be identified through the standard algebraic technique of Gram-Schmidt orthogonalization.

The general transformation from Eq.~(\ref{Eq:InterfTransformation}) taken with $\mu=1,2$ implies the following expression for the probability of a coincidence event
\begin{equation}
p_{c}(\theta)=1-\frac{1}{2}(1+{\cal V})\sin^{2}\theta,\label{Eq:pctheta}
\end{equation}
while the probability of a double event is $p_{d}(\theta)=1-p_{c}(\theta)$.
These formulas combine expressions for fully indistinguishable and completely distinguishable pairs with respective probabilities $\mathcal{V}$ and $1-\mathcal{V}$.
The resulting fringes are depicted in Fig.~\ref{Fig:Fringes}(b)
for ${\cal V}=1$ and $0.93$.

As a consequence of the Cramér-Rao bound \cite{Kay1993} the minimum
uncertainty  of any unbiased phase estimate obtained
from a measurement using $N$ photon pairs around an operating point
$\theta$ is given by
\begin{equation}
\mathit{\Delta}^{\text{pair}} = \frac{1}{\sqrt{NF^{\text{pair}}(\theta)}},
\label{Eq:Delta(2)=F2}
\end{equation}
where for standard photon counting at output ports of the interferometer
the Fisher information $F^{\text{pair}}(\theta)$ is given by a sum of
two terms contributed by coincidence and double events \cite{Demkowicz-Dobrzanski2014a},
\begin{equation}
F^{\text{pair}}(\theta)= \frac{1}{p_{c}(\theta)}\left(\frac{dp_{c}}{d\theta}\right)^{2} + \frac{1}{p_{d}(\theta)}\left(\frac{dp_{d}}{d\theta}\right)^{2} .\label{Eq:F2}
\end{equation}
As a reference, we will take the uncertainty of ideal, shot-noise-limited
phase measurement $\Delta^{\text{shot}}=1/\sqrt{2N}$, when $2N$ photons
are sent individually to the interferometer. Our figure of merit
will be the ratio $\varepsilon=\Delta^{\text{pair}}/\Delta^{\text{shot}}$ of these
two uncertainties, with $\varepsilon<1$ implying that sub-shot noise
precision has been achieved.

When the interferometer is fed with pairs of perfectly
indistinguishable photons characterized by ${\cal V}=1$, we have
$\varepsilon=1/\sqrt{2}$ independently of the operating point of
the interferometer. It can be verified that around $\theta=0$ (equivalent
to $\theta=\pi$) and $\theta=\pi/2$ the main contribution to Fisher
information defined in Eq.~(\ref{Eq:F2}) comes respectively
from double or coincidence events that occur with vanishing probabilities
when approaching these phase values. This is because in the ideal scenario
even a small number of rare events provides a sound basis to infer the phase shift.
Such a regime corresponds in standard interferometry to dark-fringe
operation used for example in gravitational wave detectors \cite{LIGO2013,Demkowicz2013}.

As seen in Fig.~\ref{Fig:Fringes}(c), the precision of phase estimation
is affected dramatically by the non-ideal indistinguishability of
photon pairs. In particular, statistical
noise generated by non-vanishing background of coincidence events effectively suppresses
information about the phase shift that could be retrieved around $\theta=\pi/2$.
We will refer to this operating point as the coincidence dark fringe.
An analogous effect would be observed also at $\theta=0$
if any mechanism generating spurious double events was incorporated
into calculations.


{\bf Restoring quantum enhancement.}
The analysis presented above assumed that we have no access to the degree of freedom introducing partial distinguishability.
For concreteness, we will consider this degree of freedom to be the spectral one, which means that all measurements performed on the photons are integrated in the frequency domain.
Suppose now that we can fully control and measure another, uncorrelated degree of freedom of the photon pairs sent to the interferometer. For the clarity of the argument, it will be convenient to use in this role the transverse spatial characteristics of the photons.
Let us consider a scenario when in addition to spectral distinguishability characterized by ${\cal V}$ we reduce the spatial overlap of the photons
 by preparing them in nonorthogonal spatial modes.
  As a result,
even in the regime of perfect spectral indistinguishability only a fraction ${\cal D}$ of photon pairs would effectively overlap in space.
To account for this scenario we will take the mode index to have two components $\mu=i\chi$, where $i=1,2$ refers to the
spectral degree of freedom, while $\chi=R,L$ denotes two mutually orthogonal spatial modes.
Using this notation, the input state $\ket{\psi_{\cal D}}$ is described by an expression analogous to Eq.~(\ref{Eq:PsiInput}) with the following substitution
of creation operators:
\begin{align}
\hat{a}_1^\dagger & \rightarrow \hat{a}_{1R}^\dagger \nonumber \\
\hat{b}_i^\dagger & \rightarrow \sqrt{\cal D} \hat{b}_{iR}^\dagger + \sqrt{1 -{\cal D}}\hat{b}_{iL}^\dagger,  \quad i=1,2
\end{align}
and it explicitly reads
\begin{multline}
\ket{\psi_{\cal D}} = \hat{a}^\dagger_{1R} \bigl[\sqrt{\cal V} \bigl( \sqrt{\cal D} \hat{b}_{1R}^\dagger + \sqrt{1 -{\cal D}}\hat{b}_{1L}^\dagger\bigr) \\
+ \sqrt{1-{\cal V}} \bigl( \sqrt{\cal D} \hat{b}_{2R}^\dagger + \sqrt{1 -{\cal D}}\hat{b}_{2L}^\dagger \bigr) \bigr] \ket{\text{vac}}
\label{Eq:psiD}
\end{multline}
Note that both the spectral components $\hat{b}_1$ and $\hat{b}_2$ have been subjected to the same spatial transformation. This is in accordance with our assumption that photon manipulations cannot depend on the inaccessible spectral degree of freedom.

\begin{figure}
\includegraphics[width=0.9\columnwidth]{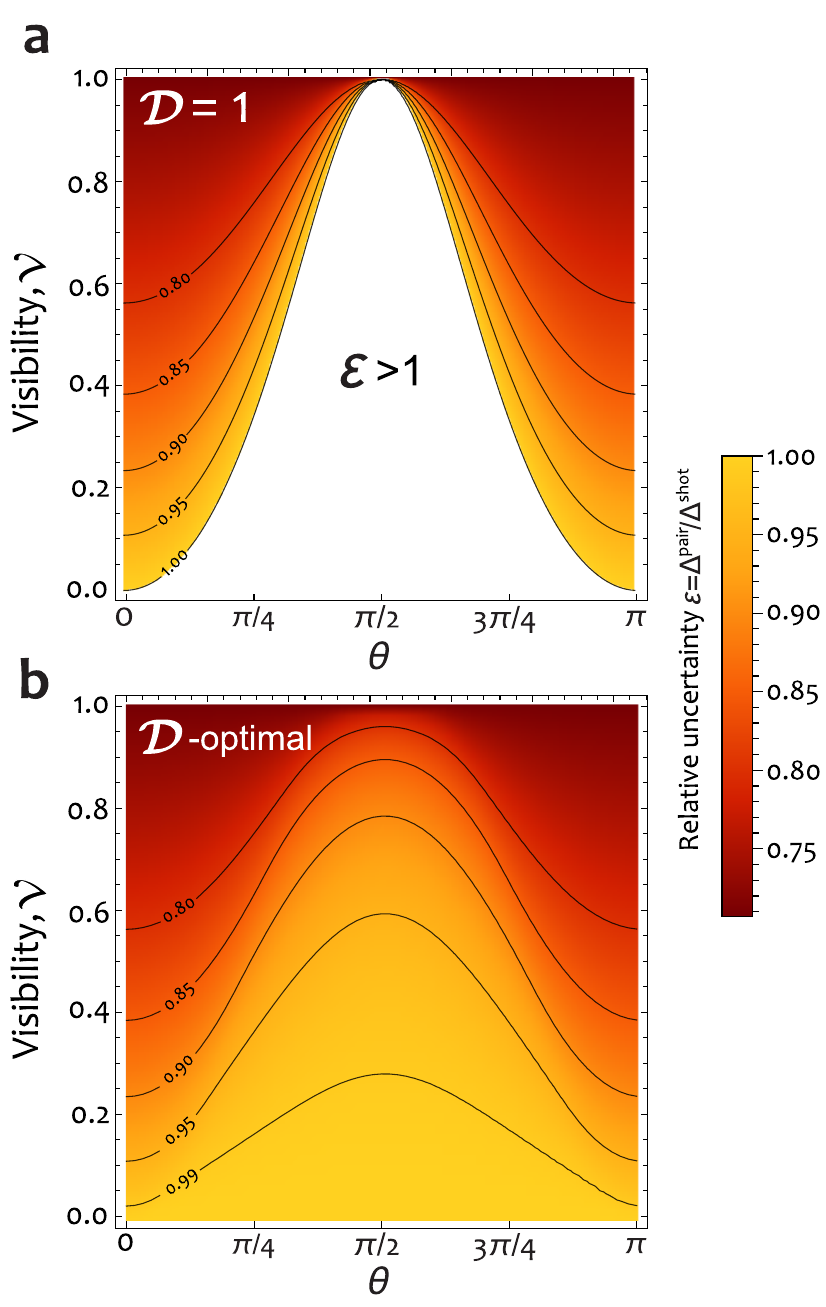}
\protect\caption{{\bf Enhancement of estimation precision.}
Relative enhancement $\varepsilon = \Delta^{\text{pair}}/ \Delta^{\text{shot}}$ for (a) full spatial overlap ${\cal D } =1$ of the two input photons and (b) the overlap parameter ${\cal D}$ optimized individually for each given spectral visibility $0 \le {\cal V} \le 1$ and an operating point $0 \le \theta \le \pi$. The white area in (a) depicts the region $\varepsilon > 1$ where sub-shot noise sensitivity is lost. It is seen that spatial mode engineering allows one to restore quantum enhancement across the entire parameter range. The uncertainty of the two-photon scheme $\Delta^{\text{pair}}$ is given by quantum Fisher information derived in Eq.~(\ref{Eq:FQthetaanyVD}).}
\label{Fig:Enhancement}
\end{figure}

Although the spectral and the spatial degrees of freedom are treated on equal footing in Eq.~(\ref{Eq:psiD}), the crucial difference is the ability to measure the latter at the interferometer output. Therefore the phase shift $\theta$ can be read out
from the paths taken by the photons as well as their transverse spatial properties.
To find the optimal strategy, we will resort to the concept of quantum Fisher information $F_Q(\theta)$ \cite{Helstrom1976,Braunstein1994}, which defines through an expression analogous to Eq.~(\ref{Eq:Delta(2)=F2}) the minimum uncertainty of a phase estimate inferred from the entire available characteristics of the physical system used for sensing. When the spatial degree of freedom for coincidence events is taken into account, the explicit expression for quantum Fisher information at the coincidence dark fringe $\theta={\pi}/{2}$, where the effects of spectral distinguishability are most severe, reads
\begin{equation}
F_Q( {\pi}/{2}) = 2 \frac{1-{\cal D}^2}{1-{\cal D}{\cal V}}.
\label{Eq:FQDarkFringe}
\end{equation}
Detailed derivation of this result is presented in Methods.
For a given $\mathcal{V}$, the maximum value of the above expression is obtained for $\mathcal{D}^{\text{opt}} = (1-\sqrt{1-\mathcal{V}^2})/\mathcal{V}$
and reads $F_Q^{\text{opt}} ({\pi}/{2}) = 4 ( 1- \sqrt{1-{\cal V}^2} )/ \mathcal{V}^2$. Remarkably, this value gives sub-shot noise precision for any $\mathcal{V} > 0$. In Fig.~\ref{Fig:Enhancement}
we compare the precision implied by numerically computed $F_Q(\theta)$ in the range $0 \le \theta \le \pi$ with $\mathcal{D}$ optimized for an individual operating point to a scenario when no mode engineering has been attempted. A cross-section of these plots for $\mathcal{V} = 0.93$ has been also shown in Fig.~\ref{Fig:Fringes}(c). It is seen that the singularity in precision around the coincidence dark fringe is removed and that the sub-shot noise operation is ensured across the entire range of $\theta$. Note that this result is achieved without any postselection of two-photon detection events and no filtering or any other manipulation in the spectral domain has been applied.

{\bf Measurement scheme.}
To elucidate the origin of the above effect it is instructive to analyze the operation of the interferometer in the $R/L$ basis of transverse spatial modes. Detecting both the photons in $RR$ modes means that they overlapped spatially at the input and therefore underwent imperfect two-photon interference affected by non-unit spectral visibility $\mathcal{V}$ with fringes shown in Fig.~\ref{Fig:Fringes}(b). On the other hand, combinations $RL$ and $LR$ at the output imply that the upper-path photon was initially in the mode $R$ and the lower-path photon in the mode $L$. Consequently, both the photons propagated through the interferometer as independent particles exhibiting single-photon interference which for $\theta=\pi/2$ gives the steepest slope of interference fringes as seen in Fig.~\ref{Fig:Fringes}(b). If the lower-path photon was prepared in a statistical mixture of $R$ and $L$ modes, only single-photon interference would provide information about the phase shift at $\theta=\pi/2$ and the shot-noise limit could not be surpassed. However, because the lower-path photon is sent into the interferometer in a superposition of the modes $R$ and $L$, information from two- and one-photon interference can be combined in a coherent way through a suitable choice of the measurement basis at the interferometer output.
Strikingly, although neither one- nor two-photon interference used separately beats the shot noise limit itself, their coherent combination restores quantum enhancement of the measurement.

\begin{figure}[b!]
\includegraphics[width=0.9\columnwidth]{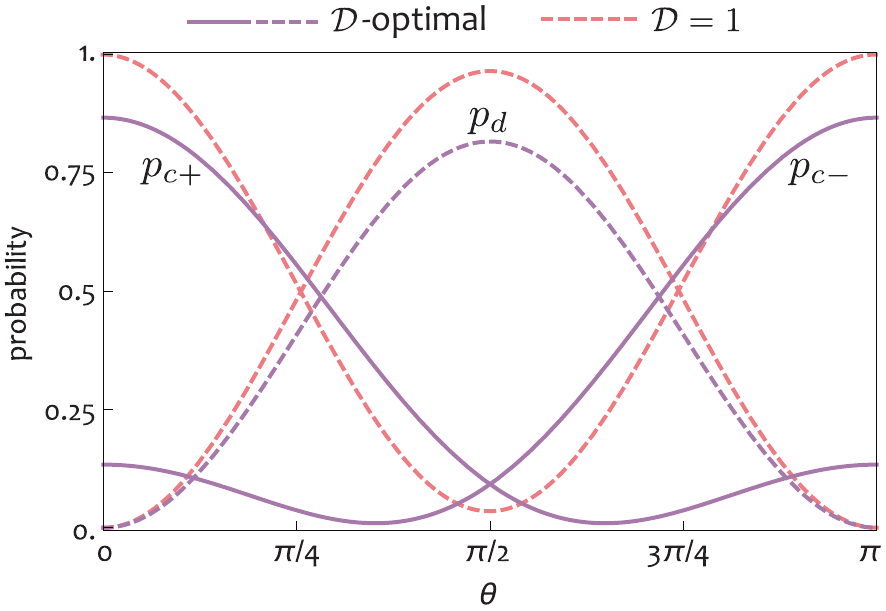}
\protect\caption{{\bf Optimized interference fringes.} The probabilities of
double $p_d$ (purple dashed line) and two types of coincidence $p_{c\pm}$ (purple solid lines) events corresponding to projections in the optimal basis $\ket{\pm}$ defined in Eq.~(\ref{Eq:ketpm}) with the overlap parameter $\mathcal{D}$ optimizing quantum Fisher information at $\theta=\pi/2$ for the visibility $\mathcal{V}=0.93$ in the inaccessible degree of freedom. Dashed red lines show standard two-photon interference fringes for the same visibility without engineering an additional degree of freedom.}
\label{Fig:ProbOpt}
\end{figure}

As derived in Methods, the explicit form of the optimal measurement attaining quantum Fisher information at the coincidence dark fringe requires discrimination between double events and two types of coincidence events corresponding to the following projections in the spatial degree of freedom:
\begin{multline}
\ket{\pm} =  \sqrt{\frac{\mathcal{D}}{1+\mathcal{D}}} \ket{RR} +
\frac{1}{2} \left( \sqrt{\frac{1- \mathcal{D}}{1+\mathcal{D}}} \pm 1 \right) \ket{RL} \\
+ \frac{1}{2} \left( \sqrt{\frac{1- \mathcal{D}}{1+\mathcal{D}}} \mp 1 \right) \ket{LR}
\label{Eq:ketpm}
\end{multline}
In Fig.~\ref{Fig:ProbOpt} we depict the resulting interference fringes for all three types of events when ${\cal D}$ is optimized for ${\cal V}=93\%$ at the operating point $\theta=\pi/2$. It is seen that coincidence events resolved in the $\pm$ basis indeed exhibit both one- and two-photon interference providing a rather steep slope at $\theta=\pi/2$, while their overall probability remains relatively low at this operating point. Combination of these features yields sub-shot noise sensitivity.

In the limit ${\cal V} \rightarrow 0 $ the optimal ${\cal D}$ approaches zero which means that at the input the two photons are nearly fully distinguishable in their spatial degree of freedom. In this regime the asymptotic expressions for the optimal measurement basis are $\ket{+} = \ket{RL}$ and $\ket{-} = \ket{LR}$, i.e.\ we need to identify the origin of a photon that has appeared at a given output port. This is equivalent to realizing single-photon interference twice, each time sending a photon into a different input port. The above analysis explains why in the limit of zero visibility ${\cal V} = 0 $ we recover the shot noise limit as evidenced by Fig.~\ref{Fig:Enhancement}(b). The described scheme is able to exploit non-classical two-photon interference for any ${\cal V} > 0 $ to achieve sub-shot noise operation, although unsurprisingly with a diminishing quantum enhancement when ${\cal V} \rightarrow 0 $.

\begin{figure*}[]
\includegraphics[width=0.7\textwidth]{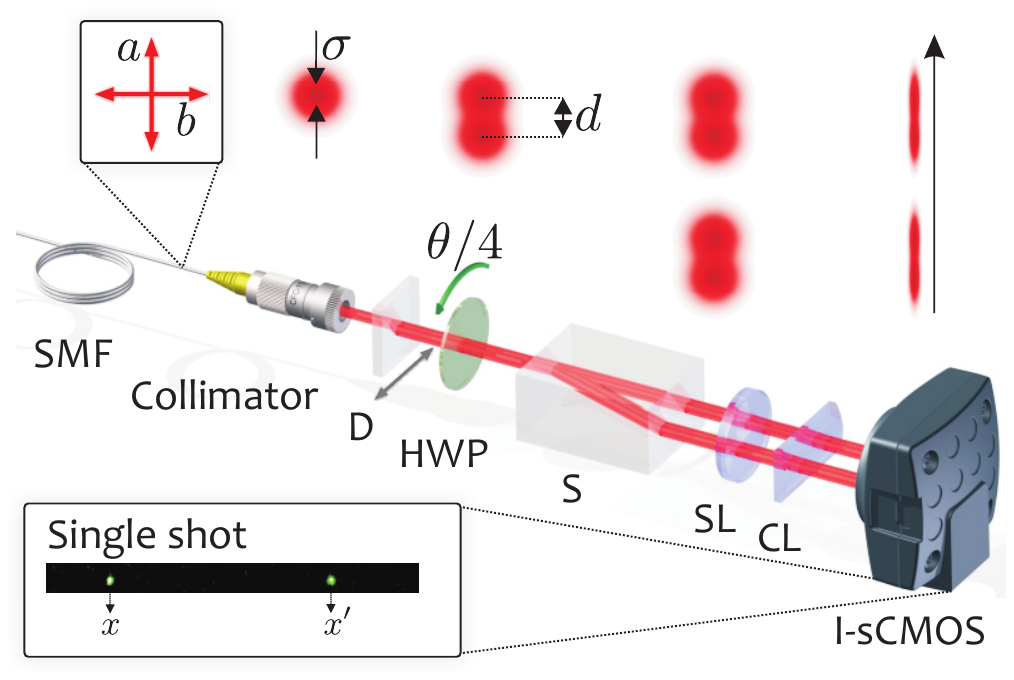}
\protect\caption{{\bf Experimental setup.} Interference between two orthogonally polarized photons $a$ and $b$ delivered by the single mode fiber SMF is realized in the common path configuration with the phase shift $\theta$ corresponding to the quadrupled rotation angle of the half-wave plate HWP.
The preceding calcite displacer D introduces transverse displacement between photon modes, which combined with spatially resolved detection restores the sub-shot-noise precision of phase estimation around the coincidence dark fringe.
The output ports of the calcite beam separator S are mapped using a spherical lens SL and a cylindrical lens CL onto the intensified sCMOS camera detecting individual coincidence events with spatial resolution as shown in the inset.  The upper right part of the figure depicts the spatial profiles of interferometer modes at the consecutive stages of the setup.}
\label{Fig:Setup}
\end{figure*}

{\bf Spatially resolved detection.}
The measurement maximizing quantum Fisher information requires implementation of rather exotic projections on two-photon superposition states given in Eq.~(\ref{Eq:ketpm}). This leads to the question about practical realization of the measurement achieving sub-shot noise precision at a given operating point. Fortunately, as we will now demonstrate, for generic spatial modes sub-shot noise precision can be  restored just by measuring transverse positions of photons emerging from the interferometer. To discuss quantitatively this idea it will be convenient to resort to the paraxial approximation and to introduce two sets of operators $\hat{a}_i(x)$ and $\hat{b}_i(x)$ labeled with a continuous one dimensional transverse position $x$. As before, the index $i=1,2$ labels spectral modes. Suppose now that the photons entering the interferometer along the upper and the lower path are prepared in spatial modes described by respective normalized profiles $u(x)$ and $v(x)$,
\begin{align}
\hat{a}_{iR} & =  \int dx \,  u(x) \hat{a}_i(x) \nonumber \\
\sqrt{\cal D} \hat{b}_{iR}^\dagger + \sqrt{1 -{\cal D}}\hat{b}_{iL}^\dagger & = \int dx \, v(x) \hat{b}_i(x).
\label{Eq:ExplicitModes}
\end{align}
The overlap parameter now reads ${\cal D} = | \int dx \, u^\ast(x) v(x) |^2$.
The probability of detecting photons in two different output ports at positions $x$ and $x'$ is given by
\begin{equation}
p_c(x,x'|\theta) = \sum_{i,j=1,2} | \bra{\text{vac}} \hat{a}_{i}(x) \hat{b}_{j}(x') \hat{U}(\theta) \ket{\psi_{\cal D}}|^2
\end{equation}
where the summation over $i=1,2$ stems from tracing over the spectral degree of freedom.

As a concrete example we will take two gaussian modes of width $\sigma$ displaced by $d$,
\begin{equation}
u(x) = \frac{1}{\sqrt[4]{2\pi\sigma^2}} e^{-(x+d/2)^2/4\sigma^2}, \quad v(x) = u(x-d)
\label{Eq:uxvx}
\end{equation}
that can be readily prepared by simple experimental means.
In the phase estimation procedure we will exploit information contained in the relative position $\xi = x-x'$ of photons detected in coincidence events. The  probability distribution for this variable is given explicitly by
\begin{multline}
p_{c}(\xi|\theta)  = \int dx'\, p_{c}(x'+\xi,x'|\theta) \\
 = \frac{1}{2\sigma\sqrt{\pi}} \left(
\cos^4 {\textstyle\frac{\theta}{2}} e^{-{(\xi-d)^{2}}/{4\sigma^{2}}} \right. \\
\left. +
\sin^4 {\textstyle\frac{\theta}{2}}
e^{-{(\xi+d)^{2}}/{4\sigma^{2}}} -
{\textstyle\frac{1}{2}}{\cal V}\sin^{2}\theta
e^{-{(\xi^{2}+d^{2})}/{4\sigma^{2}}} \right).
 \label{Eq:pc(xitheta)}
\end{multline}
When the relative position of the two photons in coincidence events is available, Fisher information defined in Eq.~(\ref{Eq:F2}) becomes enhanced by replacing the first term in the sum with the following integral over $\xi$:
\begin{equation}
\frac{1}{p_{c}(\theta)}\left(\frac{dp_{c}}{d\theta}\right)^{2}\rightarrow F_c(\theta) = \int d\xi\frac{1}{p_{c}(\xi|\theta)}\left(\frac{\partial p_{c}(\xi|\theta)}{\partial\theta}\right)^{2}.\label{Eq:XiContribution}
\end{equation}
In Fig.~\ref{Fig:Fringes}(c) we depict the estimation precision for the ratio $d/\sigma = 1.64$ which was used in the experiment described below. It is seen that the precision is brought to the sub-shot noise regime over the entire operating range.

{\bf Experiment.}
To verify experimentally sub-shot noise phase sensitivity of the interferometric scheme described above we constructed an optical setup shown in Fig.~\ref{Fig:Setup}. The interferometer is fed with 800 nm photon pairs generated via type-II spontaneous parametric down-conversion process, which are synchronized using a delayed line, spatially filtered
through a single-mode fiber and delivered to the setup in two mutually orthogonal linear polarizations corresponding to the two input ports of the Mach-Zehnder interferometer.
The photons emerging from the fiber were partly separated in space by inserting a 1.9~mm long calcite displacer D, which results in the
displacement of $d=200~\mathrm{\mu m}$ between the two ortogonally polarized output paths. The spatial modes can be modelled by gaussian functions defined in Eq.~(\ref{Eq:uxvx})
with $\sigma = 122$~$\mu$m.
To ensure temporal stability, the interferometer transformation is implemented in the
common-path configuration as a half-wave plate HWP, with the rotation angle equal to quadruple the phase shift $\theta$ between the interferometer arms, followed by a calcite beam separator S. The equivalence of this setup with the standard Mach-Zehnder interferometer is evidenced by decomposing the interferometer transformation introduced in Eq.~(\ref{Eq:InterfTransformation}) as
\begin{multline}
\begin{pmatrix}
\cos{\textstyle\frac{\theta}{2}} & \sin{\textstyle\frac{\theta}{2}}\\
\sin{\textstyle\frac{\theta}{2}} & - \cos{\textstyle\frac{\theta}{2}}
\end{pmatrix} \\
=
\begin{pmatrix}
\cos{\textstyle\frac{\theta}{4}} & -\sin{\textstyle\frac{\theta}{4}}\\
\sin{\textstyle\frac{\theta}{4}} & \cos{\textstyle\frac{\theta}{4}}
\end{pmatrix}
\begin{pmatrix}
1 & 0 \\ 0 & -1
\end{pmatrix}
\begin{pmatrix}
\cos{\textstyle\frac{\theta}{4}} & \sin{\textstyle\frac{\theta}{4}}\\
-\sin{\textstyle\frac{\theta}{4}} & \cos{\textstyle\frac{\theta}{4}}
\end{pmatrix}.
\end{multline}
In the common-path configuration the diagonal matrix with entries $1$ and $-1$ describes the half-wave plate in the coordinate system of its principal axes and the two outer matrices correspond to rotation to the laboratory reference frame by an angle $\theta/4$.

We determined through an independent measurement (see Methods) the residual spectral distinguishability of the photons to yield ${\cal V}=93\%$.
Note that this value of visibility has been used in the
example presented in Fig.~\ref{Fig:Fringes}(b). The transverse distance between
the two spatially separated modes corresponding to the output ports of the interferometer
is $3.2~\mathrm{mm}$. The rear surface of the separator S is imaged by means
of a spherical lens SL onto a single-photon-sensitive camera system (see Methods) capable of spatially resolved detection of coincidence events \cite{Chrapkiewicz2014d,Jachura2015}. The profiles of spatial modes propagating in this case through consecutive stages of the setup are shown in the upper right part
of Fig.~\ref{Fig:Setup}.

\begin{figure*}
\includegraphics[width=0.7\textwidth]{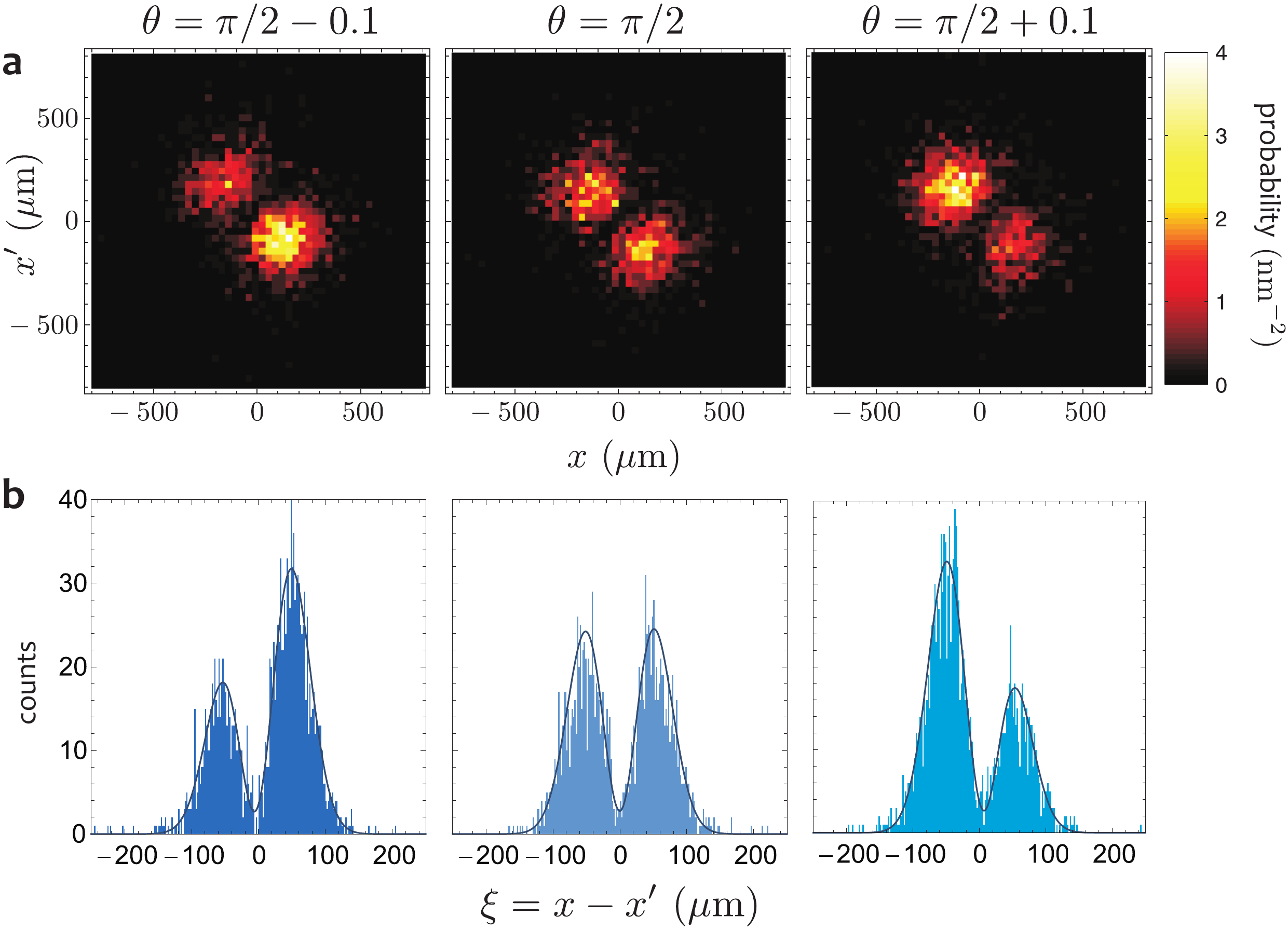}
\protect\caption{{\bf Spatial distributions of coincidence events.}
(a) Experimentally measured joint distributions $p_{c}(x,x'|\theta)$ of detecting two photons at positions $x$ and $x'$ in two different output ports of the interferometer for photon pairs characterized by spectral indistinguishability $\mathcal{V}=0.93$ and three phase shifts $ \theta = \pi/2-0.1, \pi/2, \pi/2+0.1 $. (b) Useful information about the phase shift can be extracted from the marginal histograms of the relative position between the photons $\xi = x -x'$, shown along with the fitted theoretical model $p_c(\xi|\theta)$ used in the estimation procedure (solid lines).}
\label{Fig:FistaszkiMarginals}
\end{figure*}

We recorded spatially resolved two-photon detection events for three values of the phase shift around the dark fringe, $ \theta = {\pi}/{2}-0.1, {\pi}/{2}, {\pi}/{2}+0.1$, registering approximately $6\times10^{3}$ events in each case. Experimentally observed spatial distributions $p_c(x,x'|\theta)$ of coincidence events along with their marginals $p_c(\xi|\theta)$ for $\xi = x-x'$
are presented in Fig.~\ref{Fig:FistaszkiMarginals}.
It is seen that the joint position distributions are clearly sensitive to the phase shift $\theta$,
in particular the sign of its deviation from $\pi/2$ can be unambiguously inferred from the asymmetry of the distribution with respect to the diagonal.

To quantify information about the phase shift present in spatial distributions, we performed phase estimation from the actual experimental data and determined the estimation precision. In the preliminary step, we verified the applicability of Eq.~(\ref{Eq:pc(xitheta)}) as a statistical model for the collected data assuming independently measured mode parameters $\sigma$
and $d$.
We used the maximum-likelihood method to fit ${\cal V}$ and $\theta$, obtaining
for the three cases depicted in Fig.$\,$\ref{Fig:FistaszkiMarginals}(b)
respective phase values $\theta=1.47(2), 1.57(2),
1.69(2)$  and ${\cal V}=0.93$ which is in agreement
with the visibility inferred from the independently measured Hong-Ou-Mandel dip.

In order to determine the actual estimation precision, we divided data obtained for a given HWP setting into approx.\ 600 subsets of 10 two-photon detection events and estimated the value of the phase shift separately from each subset. The width of the resulting distribution of individual estimates can be used as a figure for the estimation precision. The choice of the right estimation procedure needs some attention for small sizes of data sets. The maximum likelihood estimator is known to be asymptotically efficient \cite{Kay1993} i.e.\ it saturates the Cram{\'e}r-Rao  bound in the asymptotic limit of infinitely many independent data samples. However, its application to small data sets is not justified owing to potential biasedness. Therefore we used an estimator which is manifestly unbiased for any data size and yields the precision given by Eq.~(\ref{Eq:Delta(2)=F2})
at least in the vicinity of a given operation point. Specifically, for an experimentally measured statistical frequency distribution $f(\xi)$ of the
relative distance between the two photons, this estimator in the vicinity of $\theta_0=\pi/2$ is explicitly given by \cite{Demkowicz-Dobrzanski2014a}
\begin{equation}
\label{eq:unbiased}
\tilde{\theta}[f] = \theta_0 +
\frac{1}{F_c(\theta_0)} \displaystyle\int d\xi\,\frac{f(\xi)}{p_{c}(\xi|\theta_0)}\left.\frac{dp_{c}(\xi|\theta)}{d\theta}\right|_{\theta=\theta_0},
\end{equation}
where $F_c(\theta)$ has been defined in Eq.~(\ref{Eq:XiContribution}).
In practice, the integral in the above formula is discretized according to the width of the histogram bins. Applying this
estimator to individual data subsets yielded a distribution of phase estimates. The estimation precision was determined as the standard deviation of this distribution, shown in Fig.~\ref{Fig:Fringes}(c) for the three phase shifts along with two-sigma error bars. The obtained values, clearly situated below the shot-noise limit, demonstrate quantum enhanced operation of the interferometer despite partial spectral distinguishability of the input photons.

\section*{Discussion}

We analyzed operation of a two-photon Mach-Zehnder interferometer, which is one of the first \cite{Rarity1990} and most advanced technologically \cite{Crespi2012} examples of quantum-enhanced metrology, in a scenario which included two degrees of freedom for interfering photons. It was assumed that one degree of freedom was inaccessible experimentally. Residual distinguishability of interfering photons in this degree of freedom has a dramatically deleterious effect on the precision of phase estimation around the operating point where coincidences at the interferometer output are suppressed owing to the Hong-Ou-Mandel effect, producing a coincidence dark fringe. We showed that exploiting another, completely uncorrelated degree of freedom over which one has full experimental control can mitigate this effect, restoring quantum-enhanced precision in the entire operating range. This result is based on a rather subtle interplay between one- and two-photon interference. At the coincidence dark fringe imperfect two-photon interference alone does not provide any information about the phase shift, while one-photon interference obviously cannot surpass the shot-noise limit on its own. We demonstrated that combining coherently both types of interference through a suitably designed preparation and measurement scheme in the second degree of freedom yields precision below the shot-noise limit. The feasibility of this approach was confirmed in an experiment using the transverse position of the photons as the controllable degree of freedom which could be measured with high resolution using a single photon sensitive camera system \cite{Chrapkiewicz2014d,Jachura2015}.
The presented strategy can be also applied to other scenarios involving two independent degrees of freedom with different experimental accessibility, for example to mitigate effects of residual spatial distinguishability when only area-integrating detection is available, but photons can be suitably manipulated and measured in the spectral domain.

\begin{figure}
\includegraphics[width=0.9\columnwidth]{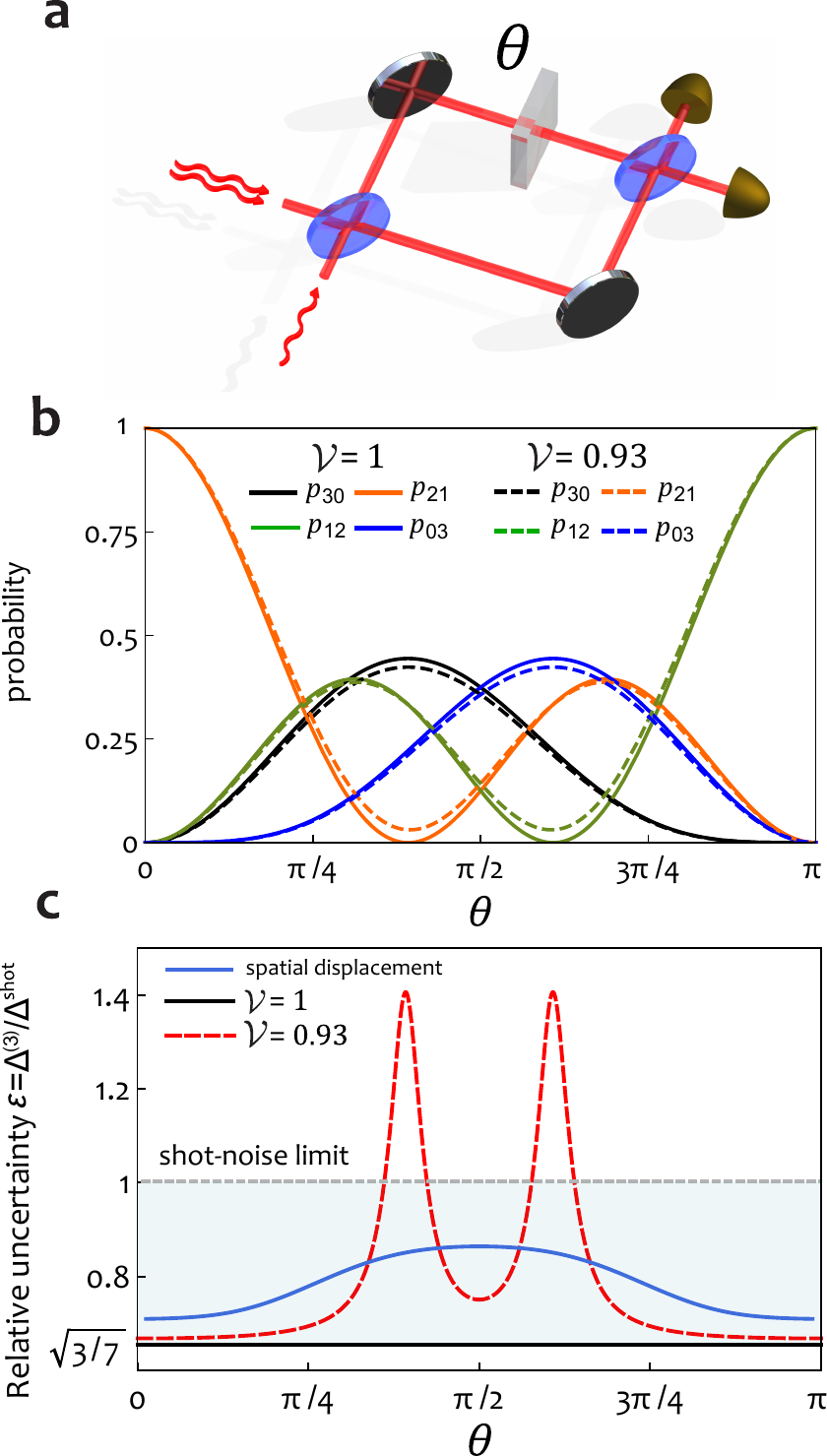}
\protect\caption{{\bf Three-photon scheme.} (a) A Mach-Zehnder interferometer fed
with a combination of two photons at one input port and a single photon at the other one.
(b) Probabilities $p_{nn'}(\theta)$ of detecting respectively $n$ and $n'$ photons at the interferometer outputs
for perfectly indistinguishable photons ${\cal V} =1$ and the single photon exhibiting residual distinguishability
with respect to the other two mutually identical photons, characterized by ${\cal V} =0.93$. (c) In the presence of residual distinguishability, the precision of phase estimation, shown with a dashed red line, deteriorates around operating points where dominant contribution to Fisher information comes from the suppression of events $nn' = 21$ or $12$. This effect can be mitigated by introducing spatial displacement $d=1.45\sigma$ between interferometer inputs, resulting in quantum enhancement across the entire operating range (solid blue line).
}
\label{Fig:2+1}
\end{figure}

Beyond two-photon interferometry, one can consider a simple scenario involving a higher photon number when the input ports of a Mach-Zenhder interferometer are fed respectively with two photons and a single photon, as shown in Fig.~\ref{Fig:2+1}(a). If all three photons are indistinguishable, counting photons at the interferometer outputs yields fringes depicted in Fig.~\ref{Fig:2+1}(b) for which Fisher information $F^{(3)}=7$ irrespectively of the operating point. This is more than two-fold improvement over the shot noise limit for three photons used independently. When the single photon exhibits residual spectral distinguishability with respect to the two photons (assumed to be identical) feeding the other port, minute changes of the shapes of the interference fringes have a dramatic effect around points $\theta_0 = 2\arctan(1/\sqrt{2})$ and $\pi-\theta_0$ on a relative phase estimation uncertainty $\varepsilon=\Delta^{(3)}/\Delta^{\textrm{shot}}$ plotted in Fig.~\ref{Fig:2+1}(c), where $\Delta^{(3)}$ is defined analogously to Eq.~\eqref{Eq:Delta(2)=F2}. According to Fig.~\ref{Fig:2+1}(b), this can be attributed to a non-vanishing background of events when two photons are detected at one output port and the remaining single photon at the other one. Preparing two photons in the same spatial mode $u(x)$ and the single photon sent individually in a partly overlapping mode $v(x)$ allows one to restore quantum enhancement as seen in Fig.~\ref{Fig:2+1}(c). Interestingly, this effect occurs at different operating points compared to the two-photon case.

A worthwhile candidate to analyze the benefits of mode engineering in a scalable multiphoton scenario may be the celebrated Holland-Burnett scheme \cite{Holland1993} employing two Fock states with equal photon numbers. It is easy to verify that sub-shot noise precision at $\theta=\pi/2$ originates from the suppression of odd photon number events at the interferometer outputs, which again is sensitive to residual distinguishability of input photons. A quantitative analysis of this scenario would require developing an efficient approach to deal with multimode multiphoton states. On the other hand, it is known that in the presence of certain common imperfections, such as photon loss, the ultimate precision follows asymptotically the shot-noise type scaling and the quantum enhancement has the form of a multiplicative factor which can be attained via a repetitive use of finite-size multiparticle superposition states \cite{Jarzyna2015a}. Therefore results obtained for a fixed number of photons may also prove useful in the asymptotic limit for realistic scenarios.

The presented results are based on the multimode description of a two-photon interferometer, which goes beyond the simplest models typically used to conceive quantum-enhanced measurement schemes. In practice, the applicability of quantum-enhanced techniques depends crucially on the ability to reduce decoherence effects caused by noise and experimental imperfections. Results presented in this paper suggest that in addition to obvious attempts to suppress decoherence effects in interferometry by improving transmission of optical elements, stabilizing phase reference, etc., exploiting the multimode structure of quantum fields can help to
to achieve the non-classical regime of operation.
If the modal structure of the probes is carefully engineered at the input and suitably detected, such a strategy can offer a noticeable improvement in precision even though decoherence effects due to another degree of freedom we do not have control over remain at the same level. An interesting question is whether analogous strategies can be generalized to quantum-enhanced interferometry with squeezed states of light \cite{LIGO2013,Demkowicz2013,Afek2010b}, boson sampling with linear multiport devices \cite{Spring2013a,Spagnolo2014,Tichy2014,Motes2014},
or perhaps benefit other quantum technologies such as optical quantum computing \cite{Knill2001,Yao2012}.


\section*{Methods}

{\bf Output two-photon state.}
For two photons prepared initially in the state $\ket{\psi_\mathcal{D}}$ defined in Eq.~(\ref{Eq:psiD}) it will be convenient to write the output state as a sum of two components,
\begin{equation}
\hat{U}(\theta) \ket{\psi_D} = \ket{\psi_c(\theta)} +  \ket{\psi_d(\theta)}
\end{equation}
Here $\ket{\psi_c(\theta)}$ is the conditional state describing coincidence events, when the two photons leave the interferometer
through different ports
\begin{multline}
\ket{\psi_c(\theta)} = - \bigl[
 \sqrt{\mathcal{D}\mathcal{V}} \hat{a}_{1R}^\dagger \hat{b}_{1R}^\dagger \cos\theta \\
 +\sqrt{\mathcal{D}(1- \mathcal{V})} \bigl( \hat{a}_{1R}^\dagger \hat{b}_{2R}^\dagger \cos^2 {\textstyle\frac{\theta}{2}}
-  \hat{a}_{2R}^\dagger \hat{b}_{1R}^\dagger \sin^2
{\textstyle\frac{\theta}{2}} \bigr) \\
 +\sqrt{(1- \mathcal{D})\mathcal{V}} \bigl( \hat{a}_{1R}^\dagger \hat{b}_{1L}^\dagger \cos^2 {\textstyle\frac{\theta}{2}}
-  \hat{a}_{1L}^\dagger \hat{b}_{1R}^\dagger \sin^2
{\textstyle\frac{\theta}{2}} \bigr) \\
 +\sqrt{(1 - \mathcal{D})(1 - \mathcal{V})} \bigl( \hat{a}_{1R}^\dagger \hat{b}_{2L}^\dagger \cos^2 {\textstyle\frac{\theta}{2}}
 - \hat{a}_{2L}^\dagger \hat{b}_{1R}^\dagger \sin^2 {\textstyle\frac{\theta}{2}}
  \bigr)\bigl] \ket{\text{vac}},
\end{multline}
while $\ket{\psi_d(\theta)}$ corresponds to double events, when both the photons emerge at the same output port of the interferometer:
\begin{multline}
\ket{\psi_d(\theta)} = {\textstyle \frac{1}{2} }\sin\theta \bigl( \sqrt{\mathcal{D}\mathcal{V}}
[(\hat{a}_{1R}^\dagger)^2 - (\hat{b}_{1R}^\dagger)^2] \\
+ \sqrt{\mathcal{D}(1-\mathcal{V})} ( \hat{a}_{1R}^\dagger \hat{a}_{2R}^\dagger - \hat{b}_{1R}^\dagger \hat{b}_{2R}^\dagger ) \\
+ \sqrt{(1-\mathcal{D})\mathcal{V}} ( \hat{a}_{1R}^\dagger \hat{a}_{1L}^\dagger - \hat{b}_{1R}^\dagger \hat{b}_{1L}^\dagger ) \\
+ \sqrt{(1-\mathcal{D})(1-\mathcal{V})} ( \hat{a}_{1R}^\dagger \hat{a}_{2L}^\dagger - \hat{b}_{1R}^\dagger \hat{b}_{2L}^\dagger )\bigl)
\ket{\text{vac}}.
\end{multline}
The overall probabilities for double and coincidence events are
\begin{equation}
p_d(\theta) = \braket{\psi_d(\theta)}{\psi_d(\theta)} = \frac{1}{2}(1+\mathcal{D}\mathcal{V})\sin^2\theta
\label{Eq:pdthetageneral}
\end{equation}
and $p_c(\theta) = 1 - p_d(\theta)$. Inserting $\mathcal{D}=1$ gives as a special case the result presented in Eq.~(\ref{Eq:pctheta}).

Let us note that all the terms in $\ket{\psi_d(\theta)}$ exhibit identical dependence on $\theta$. Consequently, resolving double events with respect to the spatial degree of freedom cannot yield more information about the phase shift. Therefore we will focus our attention on coincidence events described by $\ket{\psi_c(\theta)}$. In order to analyze information about $\theta$ when the spectral degree of freedom cannot be accessed, we will treat it formally as another subsystem $\Omega$, writing $\hat{a}^\dagger_{i \chi} \hat{b}^\dagger_{i'\chi'}\ket{\text{vac}} =  \ket{\chi\chi'} \otimes \ket[\Omega]{ii'}$, where $i,i'=1,2$ and $\chi, \chi'=R,L$. Tracing the two-photon state over the spectral subsystem yields the reduced density matrix $\hat{\varrho}_c (\theta) = \Tr_{\Omega} [ \proj {\psi_c(\theta)}]$ which written in the basis of spatial modes $\ket{RR}$, $\ket{RL}$,$\ket{LR}$, $\ket{LL}$ reads:
\begin{widetext}
\begin{equation}
\hat{\varrho}_c (\theta) =
\begin{pmatrix}
{\cal D}[1 - \frac{1}{2} (1+{\cal V}) \sin^2\theta] &
\sqrt{{\cal D}(1-{\cal D})} (\cos^4 \frac{\theta}{2} - \frac{1}{4} {\cal V} \sin^2\theta) &
\sqrt{{\cal D}(1-{\cal D})} (\sin^4 \frac{\theta}{2} - \frac{1}{4} {\cal V} \sin^2\theta) & 0  \\
\sqrt{{\cal D}(1-{\cal D})} (\cos^4 \frac{\theta}{2} - \frac{1}{4} {\cal V} \sin^2\theta) &
(1-{\cal D})\cos^4 \frac{\theta}{2} &
- \frac{1}{4}(1-{\cal D}) {\cal V} \sin^2\theta & 0 \\
\sqrt{{\cal D}(1-{\cal D})} (\sin^4 \frac{\theta}{2} - \frac{1}{4} {\cal V} \sin^2\theta) &
- \frac{1}{4}(1-{\cal D}) {\cal V} \sin^2\theta &
(1-{\cal D})\sin^4 \frac{\theta}{2} & 0 \\
0 & 0 & 0 & 0
\end{pmatrix}.
\label{Eq:varrhoctheta}
\end{equation}
\end{widetext}

{\bf Fisher information.}
If the spatial degree of freedom of the photons at the interferometer output is projected onto the $R/L$ basis, the complete statistics of measurement results is described by the diagonal elements of the density matrix $\hat{\varrho}_c (\theta)$ for coincidence events and the collective probability $p_d(\theta)$ for all double events. An easy calculation yields the corresponding Fisher information
\begin{equation}
F_{R/L}(\theta) = \frac{2(1+\mathcal{D}\mathcal{V}) - (\mathcal{D}+1)(\mathcal{V}+1)\sin^2\theta}{1- \frac{1}{2}(\mathcal{V}+1)\sin^2\theta}.
\label{Fclassical}
\end{equation}
At the dark fringe we have $F_{ R/L }({\pi}/{2})= 2(1-\mathcal{D})$. This expression, which does not even reach the shot-noise level, can be understood intuitively: information about the phase shift is obtained only from mode-mismatched pairs, when the spatial mode $R$ or $L$ at the output identifies unambiguously the input port of a given photon. The factor $1-\mathcal{D}$ in $F_{ R/L}({\pi}/{2})$ is the overall fraction of these events, while the constant $2$ is contributed by single-photon interference exhibited by such pairs.

Clearly, a measurement in the $R/L$ basis neglects information contained in the off-diagonal elements of the density matrix $\hat{\varrho}_c (\theta)$. A measurement strategy that exploits optimally $\hat{\varrho}_c (\theta)$ is described by quantum Fisher information involving the symmetric logarithmic derivative. Its calculation is simplified by switching to a new basis for the spatial degree of freedom,
\begin{align}
\ket{\alpha} & = \sqrt{\frac{2\mathcal{D}}{1+\mathcal{D}}} \ket{RR} + \sqrt{\frac{1-\mathcal{D}}{2(1+\mathcal{D})}}(\ket{RL} + \ket{LR}), \nonumber \\
\ket{\beta} & = \frac{1}{\sqrt{2}} (\ket{RL} - \ket{LR}), \nonumber
\\
\ket{\gamma} & =  \sqrt{\frac{1-\mathcal{D}}{1+\mathcal{D}}} \ket{RR} - \sqrt{\frac{\mathcal{D}}{1+\mathcal{D}}} (\ket{RL} + \ket{LR}).
\end{align}
The conditional density matrix $\hat{\varrho}_c(\theta)$ takes the form
\begin{align}
\bra{\alpha} \hat{\varrho}_c(\theta) \ket{\alpha} & = \frac{1+\mathcal{D}}{2} \left( 1- \frac{1+{\cal V}}{2} \sin^2\theta \right) \nonumber \\
\bra{\alpha}  \hat{\varrho}_c(\theta) \ket{\beta} & =
\bra{\beta} \hat{\varrho}_c(\theta) \ket{\alpha} = \frac{1}{2} \sqrt{1-\mathcal{D}^2} \cos \theta \nonumber \\
\bra{\beta}  \hat{\varrho}_c(\theta)\ket{\beta} & = \frac{1-\mathcal{D}}{2} \left( 1 - \frac{1-{\cal V}}{2} \sin^2\theta\right)
\end{align}
while all other elements involving $\ket{\gamma}$ or $\bra{\gamma}$ vanish. The symmetric logarithmic derivative $\hat{L}_c(\theta)$, given in general by the implicit formula
\begin{equation}
 \frac{d\hat{\varrho}_c}{d\theta}  =
 \frac{1}{2} [\hat{L}_c(\theta) \hat{\varrho}_c(\theta) + \hat{\varrho}_c(\theta)\hat{L}_c(\theta)]
\end{equation}
can now be easily found in the two-dimensional subspace spanned by $\ket{\alpha}$ and $\ket{\beta}$.
If double events are not resolved in the spatial degree of freedom, quantum Fisher information can be written as
\begin{multline}
F_Q(\theta) = \Tr \{ \hat{\varrho}_c(\theta)
[\hat{L}_c(\theta) ]^2\} + \frac{1}{p_{d}(\theta)}\left(\frac{dp_{d}}{d\theta}\right)^{2} \\
 = 2\frac{1-\mathcal{D}^2 + (1+\mathcal{D}\mathcal{V})^2 \cos^2\theta}%
{1- \mathcal{D}\mathcal{V} + ( 1 + \mathcal{D}\mathcal{V})\cos^2\theta}\sin^2\theta
+ 2 (1+ \mathcal{D}\mathcal{V}) \cos^2\theta,
\label{Eq:FQthetaanyVD}
\end{multline}
where $p_d(\theta)$ is given by Eq.~(\ref{Eq:pdthetageneral}). Specializing the above expression to $\theta=\pi/2$ yields Eq.~(\ref{Eq:FQDarkFringe}), whereas its value for ${\cal D}$ optimized individually for a given spectral visibility $0 \le {\cal V} \le 1$ and an operating point $0 \le \theta \le \pi$ is presented in Fig.~\ref{Fig:Enhancement}(b). The symmetric logarithmic derivative has a simple off-diagonal form at $\theta=\pi/2$
\begin{equation}
\hat{L}_c(\pi/2) = - 2 \frac{\sqrt{1-\mathcal{D}^2}}{1-\mathcal{D}\mathcal{V}} \bigl( \ket{\alpha}\bra{\beta}
+ \ket{\beta} \bra{\alpha} \bigr).
\end{equation}
Quantum Fisher information is saturated by projecting the spatial degree of freedom onto the eigenstates of $\hat{L}_c(\theta)$ \cite{Helstrom1976,Braunstein1994} given explicitly for $\theta=\pi/2$ by $\ket{\pm} = \bigl(\ket{\alpha} \pm \ket{\beta} \bigr) /\sqrt{2}$. These states are nontrivial superpositions of photon pairs prepared in combinations of $R,L$ spatial modes at two different output ports of the interferometer.

{\bf Experimental details.} In the experiment we used a photon pair source based on the II-type SPDC process in a 5-mm long periodically poled KTP crystal (Raicol Crystals) pumped with 8 mW of 400~nm light from a continuous wave diode laser. The produced pairs are transmitted through a 3~nm FWHM interference filter, carefully synchronized in time using a delay line and spatially filtered using the single mode fiber. The gaussian-like spatial modes of the photons after the fiber have a flat phase and the half-width $\sigma=122\mathrm{\,\mu m}$ at $1/e$ height for the intensity distribution measured at the position of the camera system. The residual spectral distinguishability of the photons was determined from the depth of the Hong-Ou-Mandel dip scanned using the delay line and measured  with standard avalanche photodiodes for the half-wave plate orientation corresponding to $\theta=\pi/2$.

Our camera system \cite{Jachura2015,Chrapkiewicz2014d} begins with an image intensifier (Hamamatsu V7090D) where each detected photon that induces a photoelectron emission produces a macroscopic charge avalanche resulting in a bright flash at the output phosphor screen. The flashes are subsequently imaged with a relay lens onto a fast, low-noise $6.5~\mu\mathrm{m} \times 6.5~\mu\mathrm{m}$ pixel size sCMOS sensor (Andor Zyla) and recorded as approx.\ 25-px gaussian wide spots which can be easily discriminated from the low-noise background. The central positions of the spots are retrieved from each captured frame with a subpixel resolution by a real-time software algorithm which provides full information about transverse coordinates of each registered coincidence event as illustrated in the inset of Fig. \ref{Fig:Setup}. For the sake of simplicity we consider only the coordinate $x$ in the horizontal plane of the setup and integrate the signals in the vertical direction. A cylindrical lens CL with $f\,=30\,\mathrm{{mm}}$ in front of the detector was used to reduce the vertical size of the image, producing effectively a 700 px $\times$ 22 px stripe, which significantly decreases frame readout time and allows us to reach 7 kHz collection rate of frames with exposure time $30~\mathrm{ns}$ each. The overall quantum efficiency of the camera system is 23\%.

{\bf Three-photon scheme.} A straightforward but tedious calculation shows that for a three-photon input state of the form $(\hat{a}^\dagger_{1R})^2 \bigl[\sqrt{\cal V} \bigl( \sqrt{\cal D} \hat{b}_{1R}^\dagger + \sqrt{1 -{\cal D}}\hat{b}_{1L}^\dagger\bigr)
+ \sqrt{1-{\cal V}} \bigl( \sqrt{\cal D} \hat{b}_{2R}^\dagger + \sqrt{1 -{\cal D}}\hat{b}_{2L}^\dagger \bigr) \bigr] \ket{\text{vac}}$ spatially resolved distributions for events 21 and  12  when the photons are split between the interferometer output ports into two and one are given by
\begin{multline}
  p_{21}(x_1, x_2 ; x' | \theta) = {\cal V} \left| \textstyle
  \frac{1}{2} \sin\theta \sin\frac{\theta}{2} \bigl( u(x_1) v(x_2) \right. \\
  \left. \textstyle
  + u(x_2) v(x_1) \bigr) u(x') - \cos^3 \frac{\theta}{2} u(x_1) u(x_2) v(x') \right|^2 \\
  \textstyle
  +(1-{\cal V}) \bigl[
  \frac{1}{4} \sin^2\theta \sin^2 \frac{\theta}{2} \bigl(
  |u(x_1)|^2  |v(x_2)|^2
  + |v(x_1)|^2  |u(x_2)|^2
  \bigr) \\ \times |u(x')|^2
  \textstyle
  + \cos^6 \frac{\theta}{2} |u(x_1)|^2  |u(x_2)|^2 |v(x')|^2\bigr]
\end{multline}
and
\begin{multline}
  p_{12}(x ; x_1', x_2' | \theta) = {\cal V} \left| \textstyle
  \sin^3 \frac{\theta}{2} v(x) u(x_1') u(x_2') \right.  \\
  \textstyle
  -
  \frac{1}{2} \sin\theta \cos\frac{\theta}{2} u(x) \bigl( u(x_1') v(x_2')
  \left. \textstyle
  + u(x_2') v(x_1') \bigr)   \right|^2 \\
  + \textstyle (1-{\cal V}) \bigl[
  \frac{1}{4} \sin^2\theta \cos^2 \frac{\theta}{2} |u(x)|^2 \bigl(
  |u(x_1')|^2  |v(x_2')|^2  \\
  + |u(x_2')|^2  |v(x_1')|^2
  \bigr)
  + \sin^6 \frac{\theta}{2} |v(x)|^2  |u(x_1')|^2 |u(x_2')|^2
  \bigr]
\end{multline}
where we have made use of Eq.~(\ref{Eq:ExplicitModes}). Fisher information taking into account spatially resolved detection of events 21 and 12 is given by
\begin{multline}
F^{(3)} (\theta) = \frac{1}{p_{30}(\theta)} \left( \frac{d p_{30}}{d\theta}\right) ^2 +
\frac{1}{p_{03}(\theta)} \left( \frac{d p_{03}}{d\theta}\right) ^2 \\
+ \int \frac{dx_1 \, dx_2  \, dx' }{p_{21}(x_1,x_2; x' | \theta )} \left( \frac{\partial}{\partial\theta}
p_{21}(x_1,x_2; x' | \theta )\right) ^2  \\
+ \int \frac{ dx dx_1' dx_2'}{p_{12}(x ; x_1', x_2' | \theta ) } \left( \frac{\partial}{\partial\theta}
p_{12}(x ; x_1', x_2' | \theta ) \right) ^2.
\end{multline}
Assuming Gaussian spatial modes introduced in Eq.~(\ref{Eq:uxvx}), the above expression was optimized over the displacement $d$ for $\theta_0 = 2\arctan(1/\sqrt{2})$ and ${\cal V}=93\%$. The relative uncertainty of a phase estimate for the obtained value $d=1.45\sigma$ has been depicted for the entire range $0 \le \theta \le \pi $ in Fig.~\ref{Fig:2+1}(c).

\section*{Acknowledgements}
The authors acknowledge inspiring discussion with M. G. Raymer and I. A. Walmsley.
This project was financed by the National Science Centre No. DEC-2013/09/N/ST2/02229
and by the European Commission under the FP7 projects SIQS (Grant Agreement no.\ 600645)
and PhoQuS@UW (Grant Agreement no.\ 316244) co-financed by the Polish Ministry of Science and Higher Education. R. C. was supported by Foundation for Polish Science (FNP).

\section*{Author contributions} M.J., R.C. and W.W. designed and performed the experiment. R.D.D. developed theory and analyzed experimental data. K.B. conceived the scheme and interpreted results. All authors contributed to the writing of the manuscript.


\end{document}